\documentclass[%
reprint,
amsmath,amssymb,
aps,
]{revtex4-1}
\usepackage[draft]{hyperref}
\usepackage{amsmath}    
\usepackage{graphicx}   
\usepackage{verbatim}   
\usepackage{color}      
\usepackage{subfigure}  
\usepackage{hyperref}   
\usepackage{float}
\usepackage{amsmath}

\usepackage{epsfig}

\usepackage{epstopdf}
\usepackage{upgreek}
\usepackage{wasysym}
\usepackage{siunitx}
\usepackage[utf8]{inputenc} 
\usepackage{color}
\definecolor{Gray}{gray}{0.5}
\definecolor{Red}{rgb}{1.0,0.0,0.0}
\definecolor{Blue}{rgb}{0.0,0.0,1.0}
\definecolor{Green}{rgb}{0.0,1.0,0.0}

\usepackage{tikz}
\usepackage{graphicx}
\usepackage{dcolumn}
\usepackage{bm}
\usepackage{xspace}
\usepackage{gensymb}

\usepackage{tikz}
\newcommand{\hui}[1]{#1}

\begin{document}
\preprint{APS/123-QED}

\title{Topological defect lasers}

\author{Sebastian Knitter$^1$}
\author{Seng Fatt Liew$^1$}
\author{Wen Xiong$^1$}
\author{Mikhael I. Guy$^2$}
\author{Glenn S. Solomon$^3$}
\author{Hui Cao$^1$}
\affiliation{%
$^1$Department of Applied Physics, Yale University, New Haven, Connecticut 06520, USA \\
$^2$Science \& Research Software Core, Yale University, New Haven, Connecticut 06520, USA \\
$^3$Joint Quantum Institute, NIST and University of Maryland, Gaithersburg, Maryland 20899, USA}%

\date{\today}

\begin{abstract}
We demonstrate topological defect lasers in a GaAs membrane with embedded InAs quantum dots. 
By introducing a disclination to a square-lattice of elliptical air holes, we obtain spatially confined optical resonances with high quality factor. Such resonances support powerflow vortices, and lase upon optical excitation of quantum dots, embedded in the structure. The spatially inhomogeneous variation of the unit cell orientation adds another dimension to the control of a lasing mode, enabling the manipulation of its field pattern and  energy flow landscape.

\end{abstract}
\pacs{PACS numbers: 75.60.Ej, 64.60.Ak}
\maketitle

 Topological defects have been extensively studied in nematic liquid crystals and colloids.  \cite{senyuk2013topological,muvsevivc2013nematic,loussert2013manipulating,brasselet2011electrically,brasselet2010spin,Barboza2012VortexInduction,barboza2013harnessing}.
There are various types of nematic defects that are characterized by a distinct orientational alignment of rod-shaped molecules or colloidal particles, creating a discontinuity in the director field around a fixed point. 
One important application of topological defects to photonics is the generation of optical beams with orbital angular momenta \cite{brasselet2009optical,Barboza2012VortexInduction,barboza2013harnessing}. 
Also tightly focused laser beams have been used to manipulate topological defects in liquid crystals \cite{smalyukh2007optical,brasselet2008statics,brasselet2009optical}.  
All these studies have been carried out on passive systems. 
It is not known what will happen when optical gain is introduced to the topological defect structure, whether it is possible to achieve lasing, and if so, what the lasing characteristic will be. The study, presented in this paper, \hui{aimed} to answer these questions.

In order to realize lasing in the topological defect structure, optical confinement must be strong to enhance light amplification. 
Due to the small size of liquid crystal molecules ($\sim 2~$nm) and low refractive index modulation, light cannot be effectively confined in naturally occurring topological defects. 
To improve optical confinement, we introduce topological defects to photonic crystals (PhCs).
In a PhC, periodic modulation of refractive index on the length scale of optical wavelength can produce a photonic bandgap (PBG) within which light cannot propagate \cite{soukoulis2001photonic,noda2003roadmap,sakoda2005optical,joannopoulos2011photonic}. 
Creation of a structural defect in the PhC then produces \hui{localized} states inside the PBG that \hui{typically have high quality ($Q$) factor}.  \cite{painter1999two,painter1999room,park2001nondegenerate,ryu2002square}. 
\hui{By introducing the spatial variation of the unit cell orientation, we add another degree of control of the lasing mode characteristic.} 

In this Letter, we present the design, fabrication and characterization of a topological defect laser. 
A square lattice of ellipse-shaped air holes are etched in a GaAs membrane. 
The air hole size as well as the lattice constant is on the order of the optical wavelength. 
The orientation of ellipse varies spatially to produce a topological defect. 
With the air holes at the defect center being removed, localized optical resonances were created. 
InAs quantum dots (QDs) were embedded in the GaAs membrane to provide optical gain under external excitation. 
Lasing occurred in the defect modes under optical pumping. 
Our numerical simulations reveal that the lasing modes possess optical vortices. 

\begin{figure}[H]
	\includegraphics[width=0.9\columnwidth]{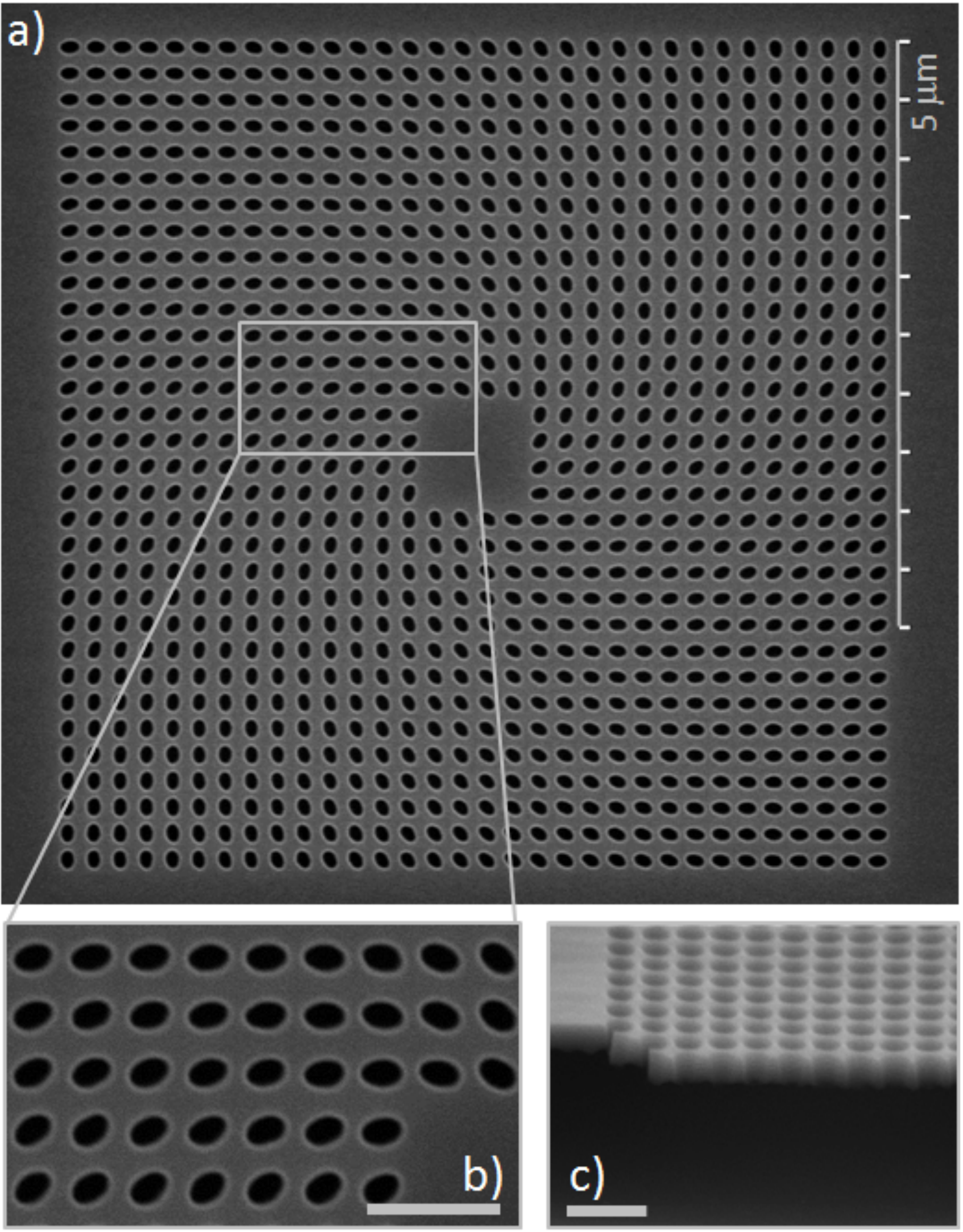}
	\caption{\label{fig:sem} Scanning-electron microscope (SEM) images of a topological defect laser fabricated in a GaAs membrane. (a) Top-view SEM image of a square lattice of $32 \times 32$ air holes with elliptical shape.  The ellipticity is $\epsilon$ = 1.4. The lattice constant is $a$ = 220 nm. The air filling fraction is 0.3. The major axis of each ellipse is rotated to an angle $\phi = \theta + \pi/4$, where $\theta$ is the polar angle of the center position of the ellipse. At the array center,  $4 \times 4$ air holes are removed. (b) Magnified SEM of a section in (a), highlighted by the gray rectangle.  (c) Tilt-view SEM image showing the free-standing GaAs-membrane. Scale-bars in (b,c) represent a length of 500 nm.}
\end{figure}

Figure~\ref{fig:sem} shows a topological defect laser that was fabricated. 
A 190-nm-thick GaAs layer and a 1000-nm thick Al$_{0.75}$Ga$_{0.25}$As layer were grown on a GaAs substrate by molecular beam epitaxy.
Inside the GaAs layer, three uncoupled layers of InAs QDs, equally spaced by 25 nm GaAs barriers, were embedded. 
The two-dimensional (2D) array of air holes was fabricated in the GaAs layer by electron-beam lithography and reactive ion etching [Fig.~\ref{fig:sem}(a)]. 
The Al$_{0.75}$Ga$_{0.25}$As layer was then etched to leave a free-standing GaAs membrane in air [Fig.~\ref{fig:sem}(c)]. 
As seen in Fig.~\ref{fig:sem}(a,b), the angle between the major axis of an ellipse and the $x$-axis was set to $\phi = k \theta + c$, where $\theta$ denotes the polar angle of the center position of the ellipse, $k=1$ is the topological charge, and $c = \pi/4$. 
$4 \times 4$ air holes were removed from the center of the topological defect, which coincides with the center of the $32 \times 32$ array, to create localized states in analogy to the regular PhC defect states. 
We have fabricated many patterns of different structural parameters, e.g., the lattice constant $a$, the filling fraction, and the ellipticity $\epsilon$  (the ratio of the major axis over the minor axis of the ellipse) of air holes. 
By changing $a$, we were able to tune the wavelength of the \hui{high-$Q$} defect-modes into the gain spectrum of InAs QDs \hui{to induce lasing action}. 
We also varied $\epsilon$ to gradually change the strength of the topological defect and monitor its effect on lasing.
In the special case of $\epsilon = 1$, the structure becomes a regular photonic crystal of circular air holes \cite{ryu2002square}. 

In the lasing experiments, the samples were optically pumped by a mode-locked Ti:Sapphire laser (pulse duration $\sim$ 200 fs,  center wavelength $\sim$ 790 nm, and pulse repetition rate $\sim$ 76 MHz). 
Optical measurements were carried out in a cryostat at $T$ = 10 K, to maximize the optical gain of InAs QDs. 
A long working distance objective lens ($50 \times$ magnification, 0.4 numerical aperture) was used to focus the pump light to the structure at normal incidence. 
The emission from the sample was collected by the same objective lens and directed to a grating (600 lines/mm) spectrometer with a cooled  charged coupled device (CCD) array detector (resolution $\sim$ 0.3 nm).

\begin{figure}[t]
	\includegraphics[width=1.0\columnwidth]{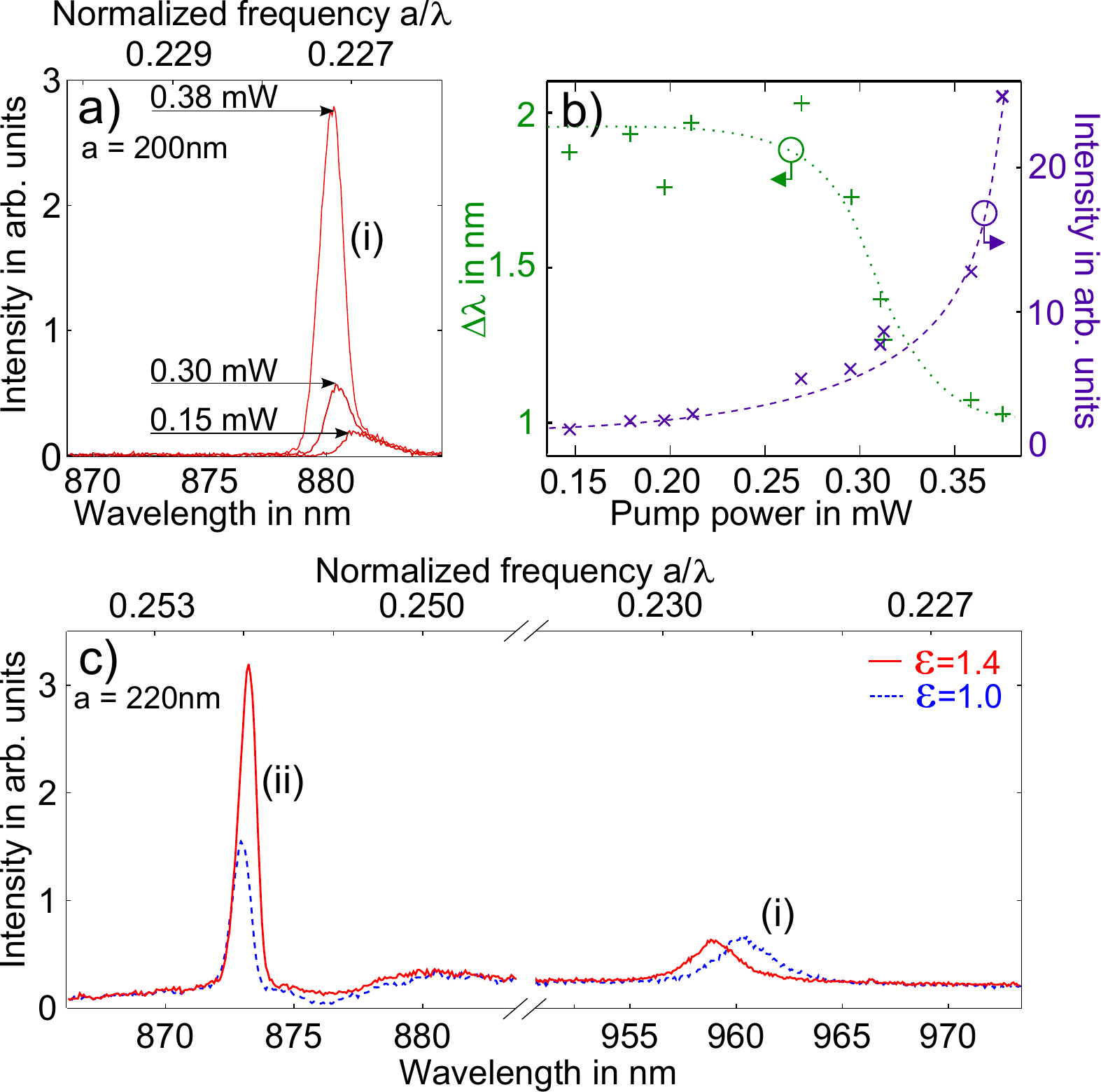}	\caption{\label{fig:spectra}
		Experimental data of lasing in the topological defect structure shown in Fig.~~\ref{fig:sem}. 
		(a) Emission spectra of the $a$ = 200 nm structure at the incident pump power of 0.15, 0.30, and 0.38 mW (from bottom to top). The emission peak is located at $\lambda$ = 880 nm ($a/\lambda$ = 0.23) and grows nonlinearly with pump power. (b) Spectrally-integrated intensity $I$ and  linewidth $\Delta \lambda$ of the emission peak in (a) as a function of the incident pump power $P$. The dotted/dashed curves guide the eye. When $P$ exceeds 0.3 mW, $I$ increases \hui{superlinearly} and $\Delta \lambda$ drops abruptly, indicating the onset of lasing action. (c) Emission spectrum of the $a$ = 220 nm structure (red solid line). The emission peak shown in (a) is shifted to 960 nm [labeled (i)] and no longer lasing. A second emission peak exhibited lasing at $\lambda$ = 873 nm ($a/\lambda$ = 0.25). For comparison, the emission spectrum of the same structure with circular holes (blue \hui{dashed} line) shows two peaks at similar wavelengths.}
\end{figure}

Lasing was observed in the topological defect structure with $a$ = 200 nm and $\epsilon = 1.4$. 
Figure~\ref{fig:spectra}(a) shows an emission peak at $\lambda = 880~$nm that grows with pump power. 
Its spectrally-integrated intensity, plotted as a function of the incident pump power $P$ in Fig.~\ref{fig:spectra}(b), exhibits a threshold behavior. 
When $P$ exceeds 0.3 mW, the peak intensity increases much more rapidly with pump power, meanwhile \hui{its linewidth} decreases abruptly. 
This behavior indicates the onset of lasing action. 
The relatively broad linewidth of $\sim$1 nm above the lasing threshold  is caused by the hot carrier effect. 
Due to short pulse pumping, the density of electron-hole pairs \hui{varies in time, causing a change in the value of refractive index and a shift of} lasing frequency \cite{pompe1995transient,jahnke1995many}. 
This transient frequency shift results in a broadening of the lasing line in the time-integrated measurement of the \hui{emission} spectrum. 

To tune the lasing frequency, we increase the lattice constant $a$ to 220 nm while keeping the filling fraction of air holes fixed. 
As shown in Fig.~\ref{fig:spectra}(c), the emission peak at 880 nm shifts to 960 nm. The normalized frequency $a/\lambda = 0.23$ remained constant, consistent with the scaling of the structure size.
Due to relatively weak gain at 960 nm, this mode [labeled (i) in Fig.~\ref{fig:spectra}(c)] no longer lased. 
\hui{Instead}, another mode [labeled (ii) in Fig.~\ref{fig:spectra}(c)] started lasing at $\lambda$ = 873 nm.
From its normalized frequency $a/\lambda = 0.25$, we infer its wavelength in the \hui{ previous structure} with $a$ = 200 nm to be $\lambda$ = 800 nm, which falls  outside the gain spectrum. Therefore this mode could not have lased previously.

We also observed lasing in the samples of $\epsilon = 1.2$ and 1.0. 
With the same lattice constant and filling fraction of air holes, the lasing frequencies are similar to those in the samples with  $\epsilon = 1.4$. 
As an example, Fig.~\ref{fig:spectra}(c) includes the emission spectrum of a sample with the same $a$ but $\epsilon$ = 1.0 (circular air holes). 
 
	\begin{figure}[bct]
		\includegraphics[width=0.9\columnwidth]{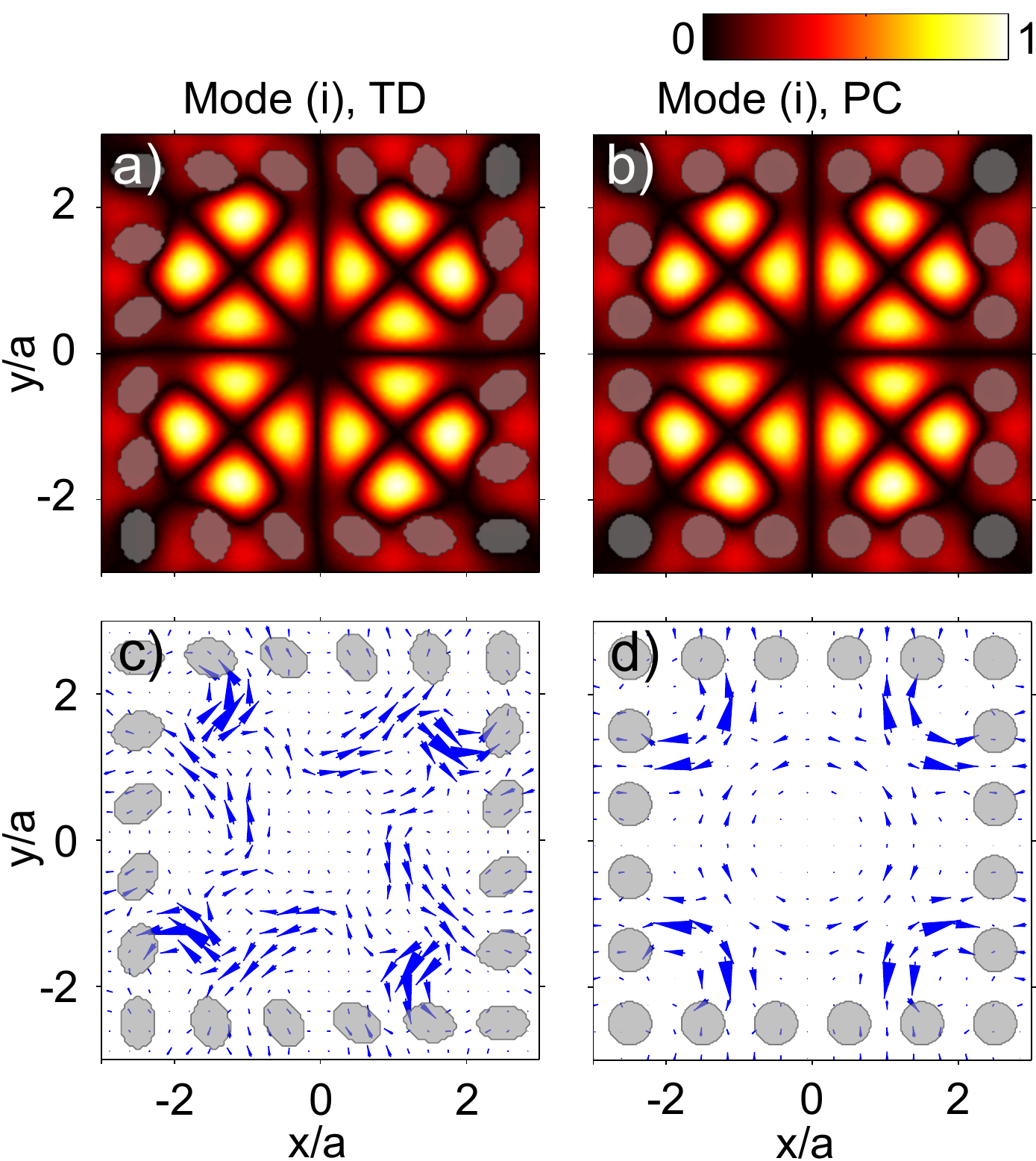}
		\caption{\label{fig:simulation1} Numerical simulation of the resonant mode in the topological defect structure (a,c) that corresponds to the lasing mode (i) in Fig.~\ref{fig:spectra}(a). For comparison, the mode in the structure with circular air holes is shown in (b,d). The air holes surrounding the central defect region are outlined in gray. The spatial distribution of the magnetic field magnitude $\left| H_z\right|$ in (a,b) reveals little difference, but the spatial map of the Poynting vector (c,d) is significantly different. Each arrow points in the direction of local energy flux, and its size is proportional to the amplitude of the flux. While the energy flows out of the central defect region in (d), it circulates clock-wise in (c).}
	\end{figure}
	
To understand the difference in the lasing modes between the topological defect structures and regular photonic crystals, we performed numerical calculations using the 3D finite-difference frequency-domain (FDFD) method \cite{comsol}. 
With the addition of optical gain, the high-$Q$ modes of the passive system, whose frequencies fall within the gain spectrum, reach the lasing threshold first, and their characteristic remains nearly unchanged above threshold. 
Therefore, the simulated resonances of the passive structures can be identified with the lasing modes that were observed experimentally.
\hui{Only transverse-electric (TE) polarized modes were computed, because the QDs in the fabricated samples provide stronger amplification for light with this polarization.}
Due to limited computing resources, the size of air hole array was reduced to $20 \times 20$.
All other structural parameters used in the simulations are identical to those of the fabricated samples.

	\begin{figure}[bct]
		\includegraphics[width=0.9\columnwidth]{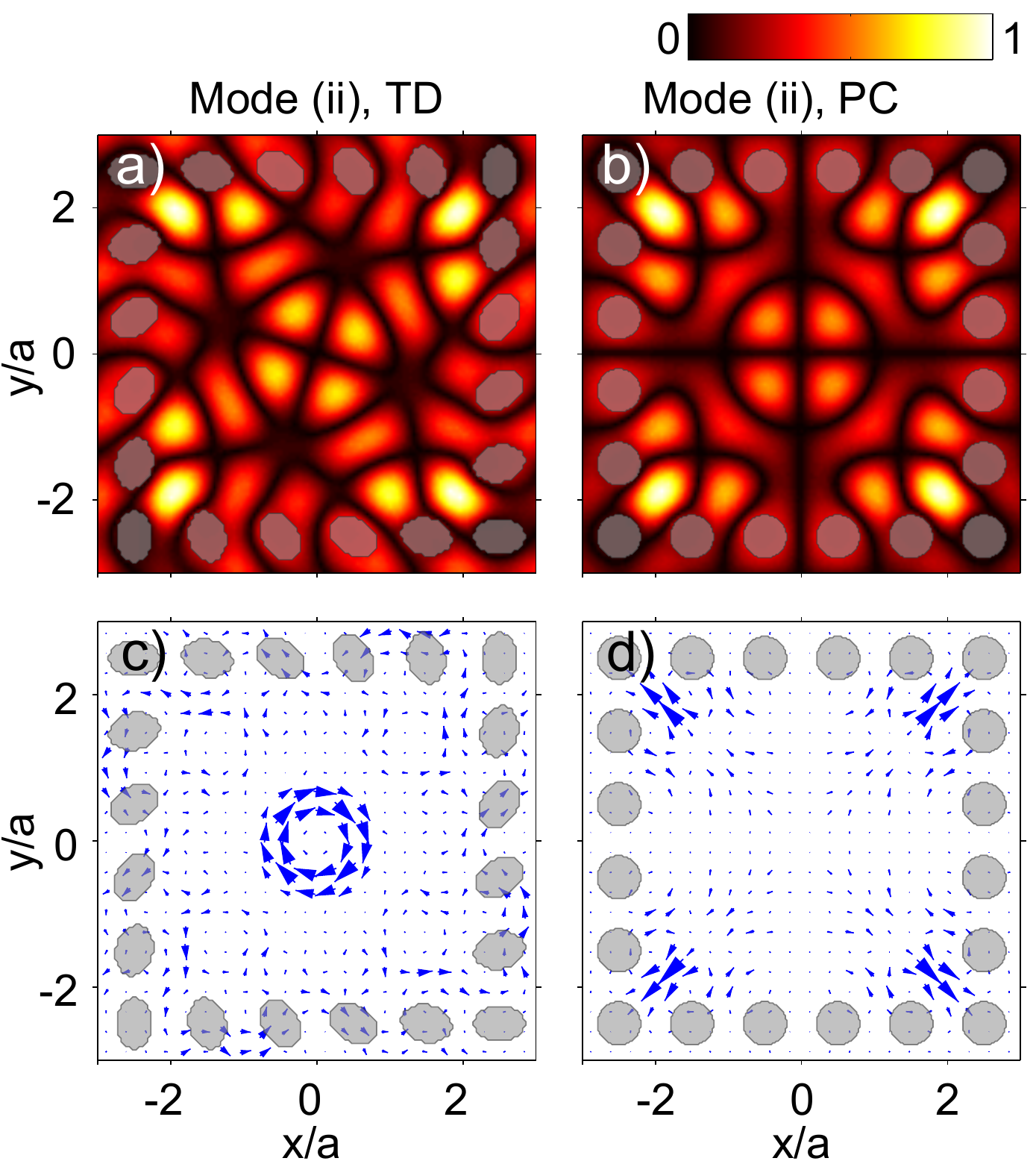}
		\caption{\label{fig:simulation2} Numerical simulation of the resonant mode in the topological defect structure (a,c) that corresponds to the lasing mode (ii) in Fig.~\ref{fig:spectra}(c). For comparison, the mode in the structure with circular air holes is shown in (b,d). The air holes surrounding the central defect region are outlined in gray. The spatial distribution of the magnetic field magnitude $\left| H_z\right|$ in (a,b) reveals the mode is rotated clockwise by the topological defect. The spatial map of the Poynting vector (c,d) illustrates the formation of an optical vortex in the topological defect structure. }
	\end{figure}
	
In the numerical simulation, we identify the high-$Q$ modes that correspond to the lasing modes observed experimentally. 
Fig. \ref{fig:simulation1}(a) shows the calculated field profile for a defect mode of normalized frequency $a/\lambda$ = 0.23, which coincides with that of the lasing mode (i) in Fig.~\ref{fig:spectra}(a).
The ellipticity of the air holes is $\epsilon$ = 1.4.
The mode quality factor is $ Q$ = \hui{$1.5\times 10^3$}.   
For comparison, we calculated the same mode in the regular PhC with circular holes ($\epsilon$ = 1.0), as shown in Fig.~\ref{fig:simulation1}(b). 
The mode profile remains nearly unchanged when the ellipticity of air holes is reduced from 1.4 to 1.0. 
However, the energy flow pattern changes significantly, as seen in the spatial map of the Poynting vector in Fig.~\ref{fig:simulation1}(c,d). 
Each arrow points in the direction of local energy flux, and its size is proportional to the amplitude of the flux.
For the regular photonic crystal defect state, the energy flows out of the central defect region [Fig.~\ref{fig:simulation1}(c)].  
In the presence of the topological defect, the optical flux circulates clockwise (CW) in the central region [Fig.~\ref{fig:simulation1}(d)]. 

Figure~\ref{fig:simulation2}(a) shows the calculated field profile for another defect mode at the normalized frequency $a/\lambda$ = 0.25, which coincides with that of the lasing mode (ii) in Fig.~\ref{fig:spectra}(c).
Unlike the mode at $a/\lambda$ = 0.23, this mode shows a notable change in the field profile when the ellipticity of the air holes is changed between 1.0 and 1.4 [\ref{fig:simulation2}(a,b)].  
More specifically, the field pattern rotates clockwise (CW) when $\epsilon$ increases from 1.0 to 1.4.  
The spatial map of the Poynting vector in Fig.~\ref{fig:simulation2}(c,d) reveals a much more dramatic change in the energy flow. 
In the PhC defect state, the energy flows mostly outward through the four corners of the defect region. 
In contrast, a CW circulating flux pattern arises in the center of a topological defect, indicating the formation of an optical vortex. 
The lateral dimension of the vortex is about \hui{one} lattice constant ($a$), which is \hui{a quarter} of the vacuum wavelength ($\lambda/4$).

\hui{The drastic change in the energy flow is attributed to the spatial variation of the ellipse orientation in the topological defect structure. 
As seen in Fig. 1(a), the topological defect structure consists of four crystalline domains located in the four quadrants. 
The ellipses in each quadrant are aligned almost in the same direction, but they are rotated 90 degree from one quadrant to the next. 
Light can leak out of the central defect region by in-plane escape through the boundary and out-of-plane scattering. 
The in-plane leakage relies on the coupling between the defect state and the propagating modes in the surrounding lattice. 
The rotation of the crystalline domains breaks the balance in the out-coupling of CW and CCW waves in the defect region. 
For example, if the CCW wave experiences more out-coupling than the CW wave, the net energy flow inside is CW \cite{liew2014photonic}.}


In summary, we experimentally realized a topological defect laser. 
By introducing the disclination to a 2D photonic crystal with anisotropic unit cell, we obtain spatially confined optical resonances with high quality factor. 
Such resonances, unlike the regular photonic crystal defect states, support powerflow vortices. 
In the presence of optical gain, these modes can lase, and their frequencies may be tuned by the structural parameters. 
This work shows that the spatially inhomogeneous variation of the unit cell orientation adds another dimension to the control of a lasing mode, enabling the manipulation of its field pattern and energy flow landscape.

We thank Yaron Bromberg, Eric Dufresne and Chinedum Osuji for useful discussions. This work is supported by the MURI grant No. N00014-13-1-0649 from the US Office of Naval Research and by the NSF under the Grant No. DMR-1205307. 

\bibliography{pubs}

\begin{thebibliography}{22}%
\makeatletter
\providecommand \@ifxundefined [1]{%
 \@ifx{#1\undefined}
}%
\providecommand \@ifnum [1]{%
 \ifnum #1\expandafter \@firstoftwo
 \else \expandafter \@secondoftwo
 \fi
}%
\providecommand \@ifx [1]{%
 \ifx #1\expandafter \@firstoftwo
 \else \expandafter \@secondoftwo
 \fi
}%
\providecommand \natexlab [1]{#1}%
\providecommand \enquote  [1]{``#1''}%
\providecommand \bibnamefont  [1]{#1}%
\providecommand \bibfnamefont [1]{#1}%
\providecommand \citenamefont [1]{#1}%
\providecommand \href@noop [0]{\@secondoftwo}%
\providecommand \href [0]{\begingroup \@sanitize@url \@href}%
\providecommand \@href[1]{\@@startlink{#1}\@@href}%
\providecommand \@@href[1]{\endgroup#1\@@endlink}%
\providecommand \@sanitize@url [0]{\catcode `\\12\catcode `\$12\catcode
  `\&12\catcode `\#12\catcode `\^12\catcode `\_12\catcode `\%12\relax}%
\providecommand \@@startlink[1]{}%
\providecommand \@@endlink[0]{}%
\providecommand \url  [0]{\begingroup\@sanitize@url \@url }%
\providecommand \@url [1]{\endgroup\@href {#1}{\urlprefix }}%
\providecommand \urlprefix  [0]{URL }%
\providecommand \Eprint [0]{\href }%
\providecommand \doibase [0]{http://dx.doi.org/}%
\providecommand \selectlanguage [0]{\@gobble}%
\providecommand \bibinfo  [0]{\@secondoftwo}%
\providecommand \bibfield  [0]{\@secondoftwo}%
\providecommand \translation [1]{[#1]}%
\providecommand \BibitemOpen [0]{}%
\providecommand \bibitemStop [0]{}%
\providecommand \bibitemNoStop [0]{.\EOS\space}%
\providecommand \EOS [0]{\spacefactor3000\relax}%
\providecommand \BibitemShut  [1]{\csname bibitem#1\endcsname}%
\let\auto@bib@innerbib\@empty
\bibitem [{\citenamefont {Senyuk}\ \emph {et~al.}(2013)\citenamefont {Senyuk},
  \citenamefont {Liu}, \citenamefont {He}, \citenamefont {Kamien},
  \citenamefont {Kusner}, \citenamefont {Lubensky},\ and\ \citenamefont
  {Smalyukh}}]{senyuk2013topological}%
  \BibitemOpen
  \bibfield  {author} {\bibinfo {author} {\bibfnamefont {B.}~\bibnamefont
  {Senyuk}}, \bibinfo {author} {\bibfnamefont {Q.}~\bibnamefont {Liu}},
  \bibinfo {author} {\bibfnamefont {S.}~\bibnamefont {He}}, \bibinfo {author}
  {\bibfnamefont {R.~D.}\ \bibnamefont {Kamien}}, \bibinfo {author}
  {\bibfnamefont {R.~B.}\ \bibnamefont {Kusner}}, \bibinfo {author}
  {\bibfnamefont {T.~C.}\ \bibnamefont {Lubensky}}, \ and\ \bibinfo {author}
  {\bibfnamefont {I.~I.}\ \bibnamefont {Smalyukh}},\ }\href@noop {} {\bibfield
  {journal} {\bibinfo  {journal} {Nature}\ }\textbf {\bibinfo {volume} {493}},\
  \bibinfo {pages} {200} (\bibinfo {year} {2013})}\BibitemShut {NoStop}%
\bibitem [{\citenamefont {Mu{\v{s}}evi{\v{c}}}(2013)}]{muvsevivc2013nematic}%
  \BibitemOpen
  \bibfield  {author} {\bibinfo {author} {\bibfnamefont {I.}~\bibnamefont
  {Mu{\v{s}}evi{\v{c}}}},\ }\href@noop {} {\bibfield  {journal} {\bibinfo
  {journal} {Phil. Trans. R. Soc. A}\ }\textbf {\bibinfo {volume} {371}},\
  \bibinfo {pages} {20120266} (\bibinfo {year} {2013})}\BibitemShut {NoStop}%
\bibitem [{\citenamefont {Loussert}\ \emph {et~al.}(2013)\citenamefont
  {Loussert}, \citenamefont {Delabre},\ and\ \citenamefont
  {Brasselet}}]{loussert2013manipulating}%
  \BibitemOpen
  \bibfield  {author} {\bibinfo {author} {\bibfnamefont {C.}~\bibnamefont
  {Loussert}}, \bibinfo {author} {\bibfnamefont {U.}~\bibnamefont {Delabre}}, \
  and\ \bibinfo {author} {\bibfnamefont {E.}~\bibnamefont {Brasselet}},\
  }\href@noop {} {\bibfield  {journal} {\bibinfo  {journal} {Phys. Rev. Lett.}\
  }\textbf {\bibinfo {volume} {111}},\ \bibinfo {pages} {037802} (\bibinfo
  {year} {2013})}\BibitemShut {NoStop}%
\bibitem [{\citenamefont {Brasselet}\ and\ \citenamefont
  {Loussert}(2011)}]{brasselet2011electrically}%
  \BibitemOpen
  \bibfield  {author} {\bibinfo {author} {\bibfnamefont {E.}~\bibnamefont
  {Brasselet}}\ and\ \bibinfo {author} {\bibfnamefont {C.}~\bibnamefont
  {Loussert}},\ }\href@noop {} {\bibfield  {journal} {\bibinfo  {journal} {Opt.
  Lett.}\ }\textbf {\bibinfo {volume} {36}},\ \bibinfo {pages} {719} (\bibinfo
  {year} {2011})}\BibitemShut {NoStop}%
\bibitem [{\citenamefont {Brasselet}(2010)}]{brasselet2010spin}%
  \BibitemOpen
  \bibfield  {author} {\bibinfo {author} {\bibfnamefont {E.}~\bibnamefont
  {Brasselet}},\ }\href@noop {} {\bibfield  {journal} {\bibinfo  {journal}
  {Phys. Rev. A}\ }\textbf {\bibinfo {volume} {82}},\ \bibinfo {pages} {063836}
  (\bibinfo {year} {2010})}\BibitemShut {NoStop}%
\bibitem [{\citenamefont {Barboza}\ \emph {et~al.}(2012)\citenamefont
  {Barboza}, \citenamefont {Bortolozzo}, \citenamefont {Assanto}, \citenamefont
  {Vidal-Henriquez}, \citenamefont {Clerc},\ and\ \citenamefont
  {Residori}}]{Barboza2012VortexInduction}%
  \BibitemOpen
  \bibfield  {author} {\bibinfo {author} {\bibfnamefont {R.}~\bibnamefont
  {Barboza}}, \bibinfo {author} {\bibfnamefont {U.}~\bibnamefont {Bortolozzo}},
  \bibinfo {author} {\bibfnamefont {G.}~\bibnamefont {Assanto}}, \bibinfo
  {author} {\bibfnamefont {E.}~\bibnamefont {Vidal-Henriquez}}, \bibinfo
  {author} {\bibfnamefont {M.~G.}\ \bibnamefont {Clerc}}, \ and\ \bibinfo
  {author} {\bibfnamefont {S.}~\bibnamefont {Residori}},\ }\href {\doibase
  10.1103/PhysRevLett.109.143901} {\bibfield  {journal} {\bibinfo  {journal}
  {Phys. Rev. Lett.}\ }\textbf {\bibinfo {volume} {109}},\ \bibinfo {pages}
  {143901} (\bibinfo {year} {2012})}\BibitemShut {NoStop}%
\bibitem [{\citenamefont {Barboza}\ \emph {et~al.}(2013)\citenamefont
  {Barboza}, \citenamefont {Bortolozzo}, \citenamefont {Assanto}, \citenamefont
  {Vidal-Henriquez}, \citenamefont {Clerc},\ and\ \citenamefont
  {Residori}}]{barboza2013harnessing}%
  \BibitemOpen
  \bibfield  {author} {\bibinfo {author} {\bibfnamefont {R.}~\bibnamefont
  {Barboza}}, \bibinfo {author} {\bibfnamefont {U.}~\bibnamefont {Bortolozzo}},
  \bibinfo {author} {\bibfnamefont {G.}~\bibnamefont {Assanto}}, \bibinfo
  {author} {\bibfnamefont {E.}~\bibnamefont {Vidal-Henriquez}}, \bibinfo
  {author} {\bibfnamefont {M.}~\bibnamefont {Clerc}}, \ and\ \bibinfo {author}
  {\bibfnamefont {S.}~\bibnamefont {Residori}},\ }\href@noop {} {\bibfield
  {journal} {\bibinfo  {journal} {Phys. Rev. Lett.}\ }\textbf {\bibinfo
  {volume} {111}},\ \bibinfo {pages} {093902} (\bibinfo {year}
  {2013})}\BibitemShut {NoStop}%
\bibitem [{\citenamefont {Brasselet}\ \emph {et~al.}(2009)\citenamefont
  {Brasselet}, \citenamefont {Murazawa}, \citenamefont {Misawa},\ and\
  \citenamefont {Juodkazis}}]{brasselet2009optical}%
  \BibitemOpen
  \bibfield  {author} {\bibinfo {author} {\bibfnamefont {E.}~\bibnamefont
  {Brasselet}}, \bibinfo {author} {\bibfnamefont {N.}~\bibnamefont {Murazawa}},
  \bibinfo {author} {\bibfnamefont {H.}~\bibnamefont {Misawa}}, \ and\ \bibinfo
  {author} {\bibfnamefont {S.}~\bibnamefont {Juodkazis}},\ }\href@noop {}
  {\bibfield  {journal} {\bibinfo  {journal} {Phys. Rev. Lett.}\ }\textbf
  {\bibinfo {volume} {103}},\ \bibinfo {pages} {103903} (\bibinfo {year}
  {2009})}\BibitemShut {NoStop}%
\bibitem [{\citenamefont {Smalyukh}\ \emph {et~al.}(2007)\citenamefont
  {Smalyukh}, \citenamefont {Kaputa}, \citenamefont {Kachynski}, \citenamefont
  {Kuzmin},\ and\ \citenamefont {Prasad}}]{smalyukh2007optical}%
  \BibitemOpen
  \bibfield  {author} {\bibinfo {author} {\bibfnamefont {I.~I.}\ \bibnamefont
  {Smalyukh}}, \bibinfo {author} {\bibfnamefont {D.~S.}\ \bibnamefont
  {Kaputa}}, \bibinfo {author} {\bibfnamefont {A.~V.}\ \bibnamefont
  {Kachynski}}, \bibinfo {author} {\bibfnamefont {A.~N.}\ \bibnamefont
  {Kuzmin}}, \ and\ \bibinfo {author} {\bibfnamefont {P.~N.}\ \bibnamefont
  {Prasad}},\ }\href@noop {} {\bibfield  {journal} {\bibinfo  {journal} {Opt.
  Expr.}\ }\textbf {\bibinfo {volume} {15}},\ \bibinfo {pages} {4359} (\bibinfo
  {year} {2007})}\BibitemShut {NoStop}%
\bibitem [{\citenamefont {Brasselet}\ \emph {et~al.}(2008)\citenamefont
  {Brasselet}, \citenamefont {Murazawa}, \citenamefont {Juodkazis},\ and\
  \citenamefont {Misawa}}]{brasselet2008statics}%
  \BibitemOpen
  \bibfield  {author} {\bibinfo {author} {\bibfnamefont {E.}~\bibnamefont
  {Brasselet}}, \bibinfo {author} {\bibfnamefont {N.}~\bibnamefont {Murazawa}},
  \bibinfo {author} {\bibfnamefont {S.}~\bibnamefont {Juodkazis}}, \ and\
  \bibinfo {author} {\bibfnamefont {H.}~\bibnamefont {Misawa}},\ }\href@noop {}
  {\bibfield  {journal} {\bibinfo  {journal} {Phys. Rev. E}\ }\textbf {\bibinfo
  {volume} {77}},\ \bibinfo {pages} {041704} (\bibinfo {year}
  {2008})}\BibitemShut {NoStop}%
\bibitem [{\citenamefont {Soukoulis}(2001)}]{soukoulis2001photonic}%
  \BibitemOpen
  \bibfield  {author} {\bibinfo {author} {\bibfnamefont {C.~M.}\ \bibnamefont
  {Soukoulis}},\ }\href@noop {} {\emph {\bibinfo {title} {Photonic crystals and
  light localization in the 21st century}}},\ Vol.\ \bibinfo {volume} {563}\
  (\bibinfo  {publisher} {Springer},\ \bibinfo {year} {2001})\BibitemShut
  {NoStop}%
\bibitem [{\citenamefont {Noda}\ and\ \citenamefont
  {Baba}(2003)}]{noda2003roadmap}%
  \BibitemOpen
  \bibfield  {author} {\bibinfo {author} {\bibfnamefont {S.}~\bibnamefont
  {Noda}}\ and\ \bibinfo {author} {\bibfnamefont {T.}~\bibnamefont {Baba}},\
  }\href@noop {} {\emph {\bibinfo {title} {Roadmap on photonic crystals}}},\
  Vol.~\bibinfo {volume} {1}\ (\bibinfo  {publisher} {Springer},\ \bibinfo
  {year} {2003})\BibitemShut {NoStop}%
\bibitem [{\citenamefont {Sakoda}(2005)}]{sakoda2005optical}%
  \BibitemOpen
  \bibfield  {author} {\bibinfo {author} {\bibfnamefont {K.}~\bibnamefont
  {Sakoda}},\ }\href@noop {} {\emph {\bibinfo {title} {Optical properties of
  photonic crystals}}},\ Vol.~\bibinfo {volume} {80}\ (\bibinfo  {publisher}
  {Springer},\ \bibinfo {year} {2005})\BibitemShut {NoStop}%
\bibitem [{\citenamefont {Joannopoulos}\ \emph {et~al.}(2011)\citenamefont
  {Joannopoulos}, \citenamefont {Johnson}, \citenamefont {Winn},\ and\
  \citenamefont {Meade}}]{joannopoulos2011photonic}%
  \BibitemOpen
  \bibfield  {author} {\bibinfo {author} {\bibfnamefont {J.~D.}\ \bibnamefont
  {Joannopoulos}}, \bibinfo {author} {\bibfnamefont {S.~G.}\ \bibnamefont
  {Johnson}}, \bibinfo {author} {\bibfnamefont {J.~N.}\ \bibnamefont {Winn}}, \
  and\ \bibinfo {author} {\bibfnamefont {R.~D.}\ \bibnamefont {Meade}},\
  }\href@noop {} {\emph {\bibinfo {title} {Photonic crystals: molding the flow
  of light}}}\ (\bibinfo  {publisher} {Princeton university press},\ \bibinfo
  {year} {2011})\BibitemShut {NoStop}%
\bibitem [{\citenamefont {Painter}\ \emph
  {et~al.}(1999{\natexlab{a}})\citenamefont {Painter}, \citenamefont {Lee},
  \citenamefont {Scherer}, \citenamefont {Yariv}, \citenamefont {O'brien},
  \citenamefont {Dapkus},\ and\ \citenamefont {Kim}}]{painter1999two}%
  \BibitemOpen
  \bibfield  {author} {\bibinfo {author} {\bibfnamefont {O.}~\bibnamefont
  {Painter}}, \bibinfo {author} {\bibfnamefont {R.}~\bibnamefont {Lee}},
  \bibinfo {author} {\bibfnamefont {A.}~\bibnamefont {Scherer}}, \bibinfo
  {author} {\bibfnamefont {A.}~\bibnamefont {Yariv}}, \bibinfo {author}
  {\bibfnamefont {J.}~\bibnamefont {O'brien}}, \bibinfo {author} {\bibfnamefont
  {P.}~\bibnamefont {Dapkus}}, \ and\ \bibinfo {author} {\bibfnamefont
  {I.}~\bibnamefont {Kim}},\ }\href@noop {} {\bibfield  {journal} {\bibinfo
  {journal} {Science}\ }\textbf {\bibinfo {volume} {284}},\ \bibinfo {pages}
  {1819} (\bibinfo {year} {1999}{\natexlab{a}})}\BibitemShut {NoStop}%
\bibitem [{\citenamefont {Painter}\ \emph
  {et~al.}(1999{\natexlab{b}})\citenamefont {Painter}, \citenamefont {Husain},
  \citenamefont {Scherer}, \citenamefont {O'Brien}, \citenamefont {Kim},\ and\
  \citenamefont {Dapkus}}]{painter1999room}%
  \BibitemOpen
  \bibfield  {author} {\bibinfo {author} {\bibfnamefont {O.}~\bibnamefont
  {Painter}}, \bibinfo {author} {\bibfnamefont {A.}~\bibnamefont {Husain}},
  \bibinfo {author} {\bibfnamefont {A.}~\bibnamefont {Scherer}}, \bibinfo
  {author} {\bibfnamefont {J.}~\bibnamefont {O'Brien}}, \bibinfo {author}
  {\bibfnamefont {I.}~\bibnamefont {Kim}}, \ and\ \bibinfo {author}
  {\bibfnamefont {P.}~\bibnamefont {Dapkus}},\ }\href@noop {} {\bibfield
  {journal} {\bibinfo  {journal} {J. Lightwave Technol.}\ }\textbf {\bibinfo
  {volume} {17}},\ \bibinfo {pages} {2082} (\bibinfo {year}
  {1999}{\natexlab{b}})}\BibitemShut {NoStop}%
\bibitem [{\citenamefont {Park}\ \emph {et~al.}(2001)\citenamefont {Park},
  \citenamefont {Hwang}, \citenamefont {Huh}, \citenamefont {Ryu},
  \citenamefont {Lee},\ and\ \citenamefont {Kim}}]{park2001nondegenerate}%
  \BibitemOpen
  \bibfield  {author} {\bibinfo {author} {\bibfnamefont {H.-G.}\ \bibnamefont
  {Park}}, \bibinfo {author} {\bibfnamefont {J.-K.}\ \bibnamefont {Hwang}},
  \bibinfo {author} {\bibfnamefont {J.}~\bibnamefont {Huh}}, \bibinfo {author}
  {\bibfnamefont {H.-Y.}\ \bibnamefont {Ryu}}, \bibinfo {author} {\bibfnamefont
  {Y.-H.}\ \bibnamefont {Lee}}, \ and\ \bibinfo {author} {\bibfnamefont
  {J.-S.}\ \bibnamefont {Kim}},\ }\href@noop {} {\bibfield  {journal} {\bibinfo
   {journal} {Appl. Phys. Lett.}\ }\textbf {\bibinfo {volume} {79}},\ \bibinfo
  {pages} {3032} (\bibinfo {year} {2001})}\BibitemShut {NoStop}%
\bibitem [{\citenamefont {Ryu}\ \emph {et~al.}(2002)\citenamefont {Ryu},
  \citenamefont {Kim}, \citenamefont {Park}, \citenamefont {Hwang},
  \citenamefont {Lee},\ and\ \citenamefont {Kim}}]{ryu2002square}%
  \BibitemOpen
  \bibfield  {author} {\bibinfo {author} {\bibfnamefont {H.-Y.}\ \bibnamefont
  {Ryu}}, \bibinfo {author} {\bibfnamefont {S.-H.}\ \bibnamefont {Kim}},
  \bibinfo {author} {\bibfnamefont {H.-G.}\ \bibnamefont {Park}}, \bibinfo
  {author} {\bibfnamefont {J.-K.}\ \bibnamefont {Hwang}}, \bibinfo {author}
  {\bibfnamefont {Y.-H.}\ \bibnamefont {Lee}}, \ and\ \bibinfo {author}
  {\bibfnamefont {J.-S.}\ \bibnamefont {Kim}},\ }\href@noop {} {\bibfield
  {journal} {\bibinfo  {journal} {Appl. Phys. Lett.}\ }\textbf {\bibinfo
  {volume} {80}},\ \bibinfo {pages} {3883} (\bibinfo {year}
  {2002})}\BibitemShut {NoStop}%
\bibitem [{\citenamefont {Pompe}\ \emph {et~al.}(1995)\citenamefont {Pompe},
  \citenamefont {Rappen}, \citenamefont {Wehner}, \citenamefont {Knop},\ and\
  \citenamefont {Wegener}}]{pompe1995transient}%
  \BibitemOpen
  \bibfield  {author} {\bibinfo {author} {\bibfnamefont {G.}~\bibnamefont
  {Pompe}}, \bibinfo {author} {\bibfnamefont {T.}~\bibnamefont {Rappen}},
  \bibinfo {author} {\bibfnamefont {M.}~\bibnamefont {Wehner}}, \bibinfo
  {author} {\bibfnamefont {F.}~\bibnamefont {Knop}}, \ and\ \bibinfo {author}
  {\bibfnamefont {M.}~\bibnamefont {Wegener}},\ }\href@noop {} {\bibfield
  {journal} {\bibinfo  {journal} {Phys. Status Solidi B}\ }\textbf {\bibinfo
  {volume} {188}},\ \bibinfo {pages} {175} (\bibinfo {year}
  {1995})}\BibitemShut {NoStop}%
\bibitem [{\citenamefont {Jahnke}\ and\ \citenamefont
  {Koch}(1995)}]{jahnke1995many}%
  \BibitemOpen
  \bibfield  {author} {\bibinfo {author} {\bibfnamefont {F.}~\bibnamefont
  {Jahnke}}\ and\ \bibinfo {author} {\bibfnamefont {S.~W.}\ \bibnamefont
  {Koch}},\ }\href@noop {} {\bibfield  {journal} {\bibinfo  {journal} {Phys.
  Rev. A}\ }\textbf {\bibinfo {volume} {52}},\ \bibinfo {pages} {1712}
  (\bibinfo {year} {1995})}\BibitemShut {NoStop}%
\bibitem [{\citenamefont {Comsol}(2014)}]{comsol}%
  \BibitemOpen
  \bibfield  {author} {\bibinfo {author} {\bibnamefont {Comsol}},\ }\href@noop
  {} {\enquote {\bibinfo {title} {Multiphysics 4.4},}\ } (\bibinfo {year}
  {2014})\BibitemShut {NoStop}%
\bibitem [{\citenamefont {Liew}\ \emph {et~al.}(2014)\citenamefont {Liew},
  \citenamefont {Knitter}, \citenamefont {Xiong},\ and\ \citenamefont
  {Cao}}]{liew2014photonic}%
  \BibitemOpen
  \bibfield  {author} {\bibinfo {author} {\bibfnamefont {S.~F.}\ \bibnamefont
  {Liew}}, \bibinfo {author} {\bibfnamefont {S.}~\bibnamefont {Knitter}},
  \bibinfo {author} {\bibfnamefont {W.}~\bibnamefont {Xiong}}, \ and\ \bibinfo
  {author} {\bibfnamefont {H.}~\bibnamefont {Cao}},\ }\href@noop {} {\bibfield
  {journal} {\bibinfo  {journal} {arXiv preprint arXiv:1411.4948}\ } (\bibinfo
  {year} {2014})}\BibitemShut {NoStop}%
\end{thebibliography}%
\end{document}